\documentclass[twocolumn,showpacs,aps,amsmath,amssymb,color,superscriptaddress]{revtex4}

\usepackage{graphicx}
\usepackage{dcolumn}
\usepackage{bm}
\usepackage[T1]{fontenc}
\begin{document}

\title{ Structure Stability and Electronic Property for Alkaline-Earth Metals Induced Si(111)-3$\times$2 Surfaces }

\author{Jun-Shuai Chai}
\affiliation{Beijing National Laboratory for Condensed Matter Physics,
             Institute of Physics, Chinese Academy of Sciences, Beijing 100190, China}
\affiliation{School of Physics, University of Chinese Academy of Sciences, Beijing 100049, China}
\author{Zhen-Zhen Li}
\affiliation{Beijing National Laboratory for Condensed Matter Physics,
             Institute of Physics, Chinese Academy of Sciences, Beijing 100190, China}
\affiliation{School of Physics, University of Chinese Academy of Sciences, Beijing 100049, China}

\author{ Lifang Xu}
\affiliation{Beijing National Laboratory for Condensed Matter Physics,
             Institute of Physics, Chinese Academy of Sciences, Beijing 100190, China}

\author{Jian-Tao Wang}
\email[e-mail address:]{wjt@aphy.iphy.ac.cn}
\affiliation{Beijing National Laboratory for Condensed Matter Physics,
             Institute of Physics, Chinese Academy of Sciences, Beijing 100190, China}
\affiliation{School of Physics, University of Chinese Academy of Sciences, Beijing 100049, China}

\date{\today}

\begin{abstract}
Structural stability and electronic properties of alkaline-earth metals (Ca, Sr, Ba) induced Si(111)-3$\times$2 surfaces have been comprehensively studied by means of {\it ab initio} calculations. Adsorption energy and charge density difference calculations show the high structural stability due to the strong chemical bonding. Analysis of electronic band structures and band-decomposed charge density distributions indicates that the third valence band is deriving from top Si and metal atoms, while the top most two valence bands are deriving from the bulk silicon.
These results suggest a larger surface band gap of 1.65$-$1.68 eV, which is good consistent with the recent experimental finding for Sr/Si(111)-3$\times$2 surface. These results reveal a natural explanation for the relevant experimental observation and stimulate further experimental and theoretical exploration on the surface science.

\end{abstract}
\maketitle

\section{Introduction}

Atoms or small clusters adsorption on the semiconductor surface are frequently reconstructed and display various exotic physical phenomena \cite{PRL2005-94-226103,PRL2006-97-046103,PRL2010-105-116102,PRL1999-82-4898}. The 3$\times$1 phase of Si(111) and Ge(111) surfaces induced by the different metal atoms [such as Ag \cite{SS1983-132-169,PRB1992-46-13635}, alkali metals (AM = Li, Na, K, Rb) \cite{SS1985-164-320,SS1988-33-38,PRB1990-41-3592,JJAP1993-32-L1263,PRB1994-49-16837} and alkaline-earth metals (AEM = Mg, Ca, Ba) \cite{JVST1986-4-1123,SS1991-355-L307,SS1996-355-271}] are the typical and well known examples. Understanding their atomic and electronic structures are very critical in the surface science and growth studies.

For the Si(111)3$\times$1-AM,Ag surfaces, the absolute coverage [1/3 monolayer (ML)] of metal atom \cite{JJAP1993-32-L1263} and the semiconductive character \cite{PRL1992-69-1419,PRB1994-50-1725,PRB1996-54-10585} were quickly determined, whereas the detail atomic geometry has a long and heavy debate. Four possible atomic models [the honeycomb chain-channel (HCC) \cite{PRL1998-81-2296,PRB1998-58-R13359} , the extended-Pandey-chain (EPC) \cite{PRL1995-75-1973}, the Seiwatz-chain \cite{PRB1996-54-8196}, and the double-$\pi$-bonded chain (D$\pi$C) \cite{PRB1998-58-3545}] have, thus, been constructed. Experimental \cite{PRL1998-80-1678,PRL1998-80-3980,SS1998-405-L503} and theoretical \cite{PRL1998-81-2296,PRB1998-58-R13359} works have confirmed that HCC model is the most promising model.

The Si(111)-3$\times$1 phase induced by AEM atoms is initially considered to have the same reconstruction as that induced by AM atoms because of the similarity of LEED $I$-$V$ cure \cite{SS1991-355-L307} for Si(111)3$\times$1-Mg and Si(111)3$\times$1-Li,Na,Ag reconstructions. The same top Si atom density (measured for Mg \cite{SS1998-415-L971}, Ca \cite{SS1999-426-298}, Na \cite{PRB1998-58-3545} and Li \cite{SS1998-405-L503}) and AEM average (measured for Ba) supported this perspective. As for the electronic properties, the AEM/Si(111)-3$\times$1 surfaces, contain an odd number of valance electron per 3$\times$1 unit cell, theoretically should be metallic different from AM/Si(111)-3$\times$1 surfaces, but it actually present the semiconducting electronic property based on the photoemission studies for Mg \cite{SSL1995-337-L789} and Ca \cite{SS2001-476-22}. For its account, an electron correlation mechanism has been introduced. In contrast to 3$\times$1 phase in LEED pattern, the scanning tunneling microscopy (STM) patterns showed a $\times$2 periodicity along the row in the empty images for Si(111)3$\times$1-Mg,Ca surfaces \cite{SS1998-415-L971,SS1999-426-298,SS2001-493-148,SS2001-476-22}, which provide a natural explanation for the semiconducting behavior with AEM average of 1/3 ML. Meanwhile, observation of half-order streaks, which is reported by Sekiguchi et al. \cite{SS2001-493-148} and Saranin et al. \cite{SS2000-448-87}, supported the possibility of 3$\times$2 phase. However, for the surface atomic structure, Lee at al. \cite{PRL2001-87-056104,PRB2003-68-115314} have studied the Ba/Si(111)-3$\times$1 LEED phase using STM and the medium-energy ion scattering (MEIS) measurements and proposal a different perspective. They find that the Ba/Si(111)-3$\times$1 surface is indeed has a 3$\times$2 reconstruction, while the Ba atom density is 1/6 ML (i.e. half of the AM coverage).
Further experimental observations also indicate their proposals \cite{PRL2001-87-056104} are universal for other AEM atoms (Ca \cite{SS2002-512-269,PRB2002-66-165319,PRB2003-68-245312,PRB2004-69-125321} and Sr \cite{APL2008-93-161912,SS2016-653-222}) In addition, the 3$\times$2 spots is hardly observed in LEED pattern for AEM/Si(111)-3$\times$2 surfaces, is primarily derive from the lack of long-range order at room temperature\cite{PRB2003-68-115314,Nature1999-402-504,PRL2000-85-808}.

On the theoretical side, for atomic structure, pervious works \cite{PRL2001-87-056104,PRB2003-68-115314,SS2006-600-3606} on AEM adatoms on Si(111)-3$\times$2 surface have mainly focus on the energetics and confirmed that the AEM/Si(111)-3$\times$2 surfaces is indeed consisted of HCC model structure and a AEM average of 1/6 ML based on the analysis of theoretical STM images. This atomic structure is also suitable for the case of Sm \cite{PRB2003-67-195413,PRB2007-75-205420} and Yb \cite{SS2010-604-1899} adatoms on Si(111)-3$\times$2 surface. For the electronic properties, previous works have studied the surface states dispersion for Si(111)-3$\times$2-Ba,Ca surfaces\cite{PRB2003-67-085411,PRB2005-72-085325}, which is similar to the case of Li/Si(111)-3$\times$1 surface, in consistent with the experimental findings \cite{PRB2003-68-245312,PRB2003-67-085411}. However, a comprehensive study on structural stability from kinetics and the surface band gap for this system have yet to be explored.

In this paper, we present a systematically investigation of the structural stability and electronic properties of AEM (Ca, Sr, Ba) induced Si(111)-3$\times$2 surface by means of {\it ab initio} calculations. Adsorption energy and charge density difference calculations show the high structural stability due to the strong chemical bonding. Analysis of electronic band structures and band-decomposed charge density distributions indicates that the first conduction and the third valence are derived from the top Si and metal atoms, while the first and second valence bands are derived from the bulk Si atoms. As a result, a larger surface band gap of 1.65$-$1.68 eV is obtained, in good consistent with the experimental observation for Sr/Si(111)-3$\times$2 surface \cite{SS2016-653-222}.

\begin{table}
\caption{Calculated adsorption energy ($E_{ad}$, in eV/atom) for AEM (Ca, Sr, Ba) adatoms on Si(111)-3$\times$1 and Si(111)-3$\times$2 surfaces with various adsorption sites.} 
\begin{tabular}{ m{2.4cm} m{1.3cm}<{\centering}   m{1.3cm}<{\centering}  m{1.3cm}<{\centering} m{1.2cm}<{\centering}}
 \hline
 \hline
Structure&$T_{4}$&$H_{3}$&$B_{2}$&$C_{6}$\\
 \hline
Ca/Si(111)-3$\times$2&-4.73&-4.65&-4.42&-3.62\\
Ca/Si(111)-3$\times$1&-3.64&-3.57&&\\
Sr/Si(111)-3$\times$2&-4.54&-4.47&-4.26&-3.44\\
Sr/Si(111)-3$\times$1&-3.22&-3.19&&\\
Ba/Si(111)-3$\times$2&-4.93&-4.89&-4.71&-4.05\\
Ba/Si(111)-3$\times$1&-3.43&-3.40&&\\

 \hline
\end{tabular}
\end{table}

\begin{figure}
\includegraphics[width=7.5cm]{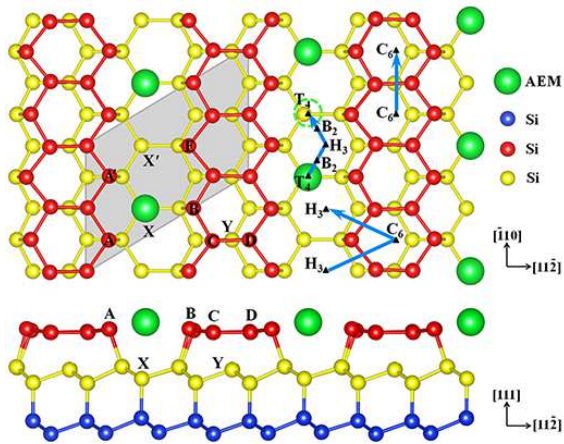}
\caption{(Color online) The Schematic depiction of AEM/Si(111)-3$\times$2 structures. The top first and second layer Si atoms are marked with red and yellow, respectively, and the green atoms are represent AEM atoms.}
\end{figure}

\section{Computational method}

The reported {\it ab initio} total energy calculations are performed using the Vienna {\it ab initio} simulation package (VASP) code \cite{vasp}, employing a plane wave basis set with 500 eV cutoff and the generalized gradient approximation using the PBE functional \cite{pbe}. The projector augmented wave (PAW) method \cite{paw} is adapted to describe the electron-ion with 3$s^{2}$3$p^{2}$ for Si, 3$s^{2}$3$p^{6}$4$s^{2}$ for Ca, 4$s^{2}$4$p^{6}$5$s^{2}$ for Sr and 5$s^{2}$5$p^{6}$6$s^{2}$ for Ba treated as valence electrons. The Brillouin zone is sampled using the Monkhorst-Pack scheme \cite{PRB1992-45-13244}. The supercell size is set to $XYZ$ = 11.60 {\AA}$\times$7.73 {\AA}$\times$30.21 {\AA} with five layers of silicon, one layer of hydrogen to passivate the lowest Si layer, and a vacuum layer of about 18 {\AA} in the $Z$ direction. The $XY$ plane corresponds to 3$\times$2 slab with periodic boundary conditions. Throughout the calculations, only one bottom Si layer is fixed at the bulk structure, while the other atoms are fully relax. The energy minimization is done over the atomic and electronic degrees of freedom using the conjugate gradient iterative technique. The convergence criteria of electronic self-consistent is set to 10$^{-5}$ eV for total energy.

\section{Results and discussion}
\subsection{Structural stability}

There are four possible adsorption sites ($T_{4}$, $H_{3}$, $B_{2}$ and $C_{6}$) for AEM adatoms (see Fig. 1). The corresponding adsorption energies listed in Table I are defined as
\begin{equation}
E_{ad}=E[AEM/Si(111)]-E[Si(111)]-E[AEM]
\end{equation}
where $E[AEM/Si(111)]$, $E[Si(111)]$ and E[AEM] are the total energy of the AEM/Si(111)-3$\times$2 system, the clean Si(111)-3$\times$2 surface, and one isolated AEM atoms, respectively. Site $T_{4}$ in the channel is the most stable adsortption site, the same as for Yb on the Ge(Si)(111)-3$\times$2 surface \cite{SS2010-604-1899} and Sm on the Si(111)-3$\times$2 surfaces \cite{PRB2003-67-195413,PRB2007-75-205420}; site $H_{3}$ in the channel is the second most stable adsorption site, which is only about 0.04$\sim$0.08 eV/AEM higher than site $T_{4}$ in energy. A top adsorption site $C_{6}$, located about 1.78$\sim$2.28 {\AA} above the honeycomb chain, is the most unstable adsorption site with about 1.04$\sim$1.11 eV energy loss than $T_{4}$, which imply that a weak interaction exist between AEM adatoms and honeyconb chain due to the formation of big $\pi$ bond.
On the other hand, for an AEM average of 1/6 (ML), we can explain this fact from two aspects: (i) the adsorption energies of $T_{4}$ and $H_{3}$ sites for AEM/Si(111)-3$\times$1 surfaces are higher by 1.09$\sim$1.50 and 1.08$\sim$1.49 eV/AEM than that for AEM/Si(111)-3$\times$2 surfaces, respectively, which suggest that AEM adatoms can not entirely fill the $T_{4}$ (or $H_{3}$) sites; (ii) the distance (3.87 {\AA}) between two neighboring AEM atoms in AEM/Si(111)-3$\times$1 surfaces is smaller than the bond length (4.08$\sim$4.87 {\AA}) of isolate AEM dimer, which indicate that a strong repulsion exist between two neighboring AEM atoms, in agreement with previous theoretical work \cite{PRB2003-68-115314}.

The optimized atomic structure for AEM/Si(111)-3$\times$2 surfaces with AEM atoms at $T_{4}$ site, as shown in Fig. 1, can be described by two parts: (i) nearly planar honeycomb chain of Si separated by empty channels in the topmost layer and (ii) AEM atoms in the channel with a average of 1/6 ML. The honeycomb chain is formed by five inequivalent Si surface atoms (labeled A, B, C, D, E) which are threefold coordinated. The inner Si(C) and Si(D) atoms are only very weakly bonded to the second layer Si(Y) atoms \cite{PRL1998-81-2296,PRB1998-58-R13359}, which allowing them undergo a favorable rehybridization from $sp^{3}$ to $sp^{2}$ and $p^{z}$, and then two released electrons favorably fill the $\pi$ bond state and form a more stable Si double bond between Si(C) and Si(D) atoms, similar to AM/Si(111)-3$\times$1 surfaces \cite{PRB2001-63-085323,SS2000-463-183}. The AEM atoms is located at position of nearly equal distance from top-layer Si(A, A$^{\prime}$ and B) atoms.

We next examine the diffusion of AEM adatoms on the Si(111)-3$\times$2 surface to study the structural stability. According to the energetic analysis above (see Table I), three possible diffusion pathway (see Fig. 1) are considered: $P_{1}$ ($C_{6}$-$C_{6}$) is a rolling-over pathway, on the honeycomb chain along the [\={1}10] direction; $P_{2}$ ($T_{4}$-$H_{3}$-$T_{4}$) is a translation pathway, along empty channel, and $P_{3}$ ($H_{3}$-$C_{6}$-$H_{3}$) is a jumping pathway, across the honeycomb chain. For AEM adatoms diffusion along path $P_{1}$($P_{2})$ and $P_{3}$,  only the $y$ and $x$ directions of AEM atoms are fixed, respectively, while all the other atoms are fully relaxed expect for the lowest Si atom layer.

The diffusion of AEM adatoms on the top of honeycomb chain is characterized by an energy barrier of 0.85$\sim$0.97 eV along the pathway $P_{1}$. The high energy barrier is due to the high structure stability of  HCC model. Meanwhile, the diffusion of AEM adatoms along the empty channel (pathway $P_{2}$) has a lower energy barrier (0.22$\sim$0.31 eV), which indicate that the AEM atoms can easily diffuse via site $H_{3}$ along the channel. The low energy barrier is due to the high concentration of Si dangling bonds in the empty channel. Moreover, a higher energy barrier (1.58$\sim$2.05 eV) for AEM adatoms diffusion along pathway $P_{3}$ imply that AEM adatoms in the channel is very hard to cross and jump to the top of honeycomb chain. Thus, pathway $P_{2}$ along the empty channel is the most likely diffusion pathway for AEM adatoms on Si(111)-3$\times$2 surface.

\begin{figure}
\includegraphics[width=7.5cm]{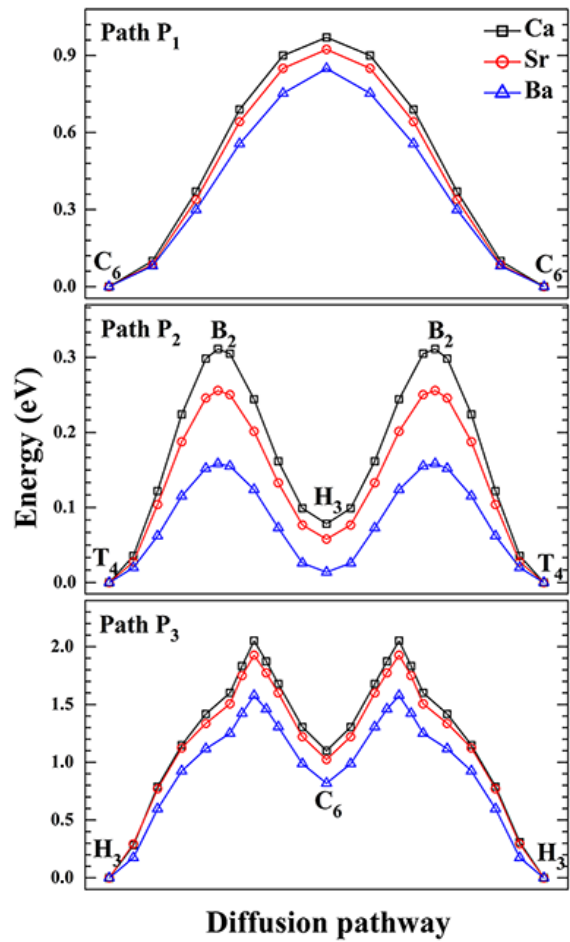}
\caption{(Color online) Relative total energy versus the different diffusion pathways for AEM adatoms on Si(111)-3$\times$2 surface. $P_{1}$ ($C_{6}$-$C_{6}$) is a rolling-over pathway, along the honeycomb chain; $P_{2}$ ($T_{4}$-$H_{3}$-$T_{4}$) is a translation pathway, along the empty channel, and $P_{3}$ ($H_{3}$-$C_{6}$-$H_{3}$) is a jump pathway, across the honeycomb chain. The typical atomic configure are shown in Fig.1}
\end{figure}

\begin{table*}
\caption{Electronic charge transfers for AEM/Si(111)-3$\times$2 structures from Mulliken analysis. Positive and negative values correspond to the loss and gain of electrons, respectively. The indicated values are in electrons/atom.}
\begin{tabular}{m{2.8cm} m{1.3cm}<{\centering} m{1cm}<{\centering}  m{1cm}<{\centering} m{1cm}<{\centering} m{1cm}<{\centering} m{1cm}<{\centering} m{1cm}<{\centering} m{1cm}<{\centering} m{1cm}<{\centering} m{1cm}<{\centering} }
 \hline
 \hline
            structure&AEM  &A    &A$^{\prime}$&    B&C    &D    &E    &X     &X$^{\prime}$   &Y\\
 \hline
Ca/Si(111)-3$\times$2&+1.68&-0.39&   -0.39   &-0.42&+0.05&+0.06&-0.17&-0.15&-0.01&-0.05\\
Sr/Si(111)-3$\times$2&+1.15&-0.28&   -0.29   &-0.30&+0.05&+0.05&-0.18&-0.09&0.00&-0.05\\
Ba/Si(111)-3$\times$2&+1.10&-0.26&   -0.27   &-0.29&+0.05&+0.04&-0.21&-0.07&0.00&-0.04\\
 \hline
 \hline
\end{tabular}
\end{table*}

For the pathway $P_{2}$, we find the AEM adatoms to move along the empty channel, it must overcome the barrier energy between $T_{4}$ and $H_{3}$ sites, and the energy barrier site is located at site $B_{2}$. The corresponding energy barrier are estimated to be 0.31, 0.28 and 0.22 eV for Ca, Sr and Ba on Si(111)-3$\times$2 surface, respectively, which is primarily attributed to charge transfer between AEM adatoms and the Si substrate, and the corresponding discussion is provided blow.
The barrier for Ca and Sr are similar because of their similar atomic size, whereas the barrier for Ba with a larger atomic size is lower than Ca and Sr. It is interesting to note that similar phenomena have also been obtained in other systems  \cite{PRL2005-94-226103}. For example, for As, Sb and Bi adatoms on Si(100) surface, Wang et al. \cite{PRL2005-94-226103} found that the energy barriers gradually increase (1.13$\sim$1.19$\sim$1.31 eV) in the order of Bi, Sb and As, and the value for Bi and Sb with similar atomic size are close. Compared to As, Sb and Bi adatoms on Si(100) surface, our case of AEM adatoms on Si(111)-3$\times$2 surface is obviously a easier diffusion process. In addition, a recent study using STM and scanning tunneling spectroscopy observed a concerted motion of Sr adatom along the empty channel on Si(111)-3$\times$2 surface at room temperature, supporting our calculations \cite{SS2016-653-222}.

\subsection{Chemical bonding}

To understand the the nature of chemical bonding for AEM adatoms on Si(111)-3$\times$2 surface, we calculate the charge density difference. It is defined as
\begin{equation}
\Delta\rho(r)=\rho[AEM/Si(111)]-\rho[Si(111)]-\rho[AEM]
\end{equation}
where first term is the charge density of AEM/S(111)-3$\times$2 system, and the second term is that of the corresponding clean Si(111)-3$\times$2 surface, and $\rho[AEM]$ is the charge density of one isolated AEM adatoms. The charge density difference for AEM adatoms on Si(111)-3$\times$2 surface have been plotted in Fig. 3, and solid and dashed lines represent the gain and loss of electrons, respectively. The results show that a large charge accumulation exist between AEM and the nearest-neighbor Si(A, A$^{\prime}$ and B) atoms, and correspondingly, a remarkably charge depletion also exist around the AEM adatoms. This suggest that there are certain charge transfer from AEM adatoms to the surface Si(A, A$^{\prime}$ and B) atoms. Hence, the bonding of AEM adatoms on Si(111)-3$\times$2 surfaces are ionic, and the major factor governing its equilibrium will be the Coulomb interaction with the saturated dangling-bond states of Si(A, A$^{\prime}$ and B) atoms \cite{PRB1998-58-R13359}. A relative smaller charge accumulation near Si(X) (just blow AEM atoms) suggest the existence of a weak Coulomb interaction between AEM and Si(X) atoms. This finding is due to the fact that the distances (3.23 {\AA} for Ca, 3.43 {\AA} for Sr and 3.66 {\AA} for Ba) between AEM and Si(X) is larger than the distances (about 2.92 {\AA} for Ca, 3.01 {\AA} for Sr, and 3.11 {\AA} for Ba) between AEM and Si(A, A$^{\prime}$, and B). Similarly, a larger distances (4.78 {\AA} for Ca, 4.76 {\AA} for Sr and 4.75 {\AA} for Ba) between AEM and Si(E) atoms can also explain a small charge accumulation near Si(E) atom.

Table II present the electronic charge transfer for AEM/Si(111)-3$\times$2 surfaces based on the Mulliken population analysis. We find that the effective charges for Ca, Sr and Ba adatoms are +1.68$e$, +1.15$e$ and +1.10$e$, respectively, which are larger than the value (+0.72$e$) for Ca reported by Miwa \cite{PRB2005-72-085325}. Almost all the charge of AEM adatoms is transfer to the nearest-neighbor surface Si(A, A$^{\prime}$ and B) atoms. It is interesting to note that the change trend of charge transfer for AEM adatoms is in contrast to that of element electronegativity, which provides a natural explanation for the change of energy barrier for AEM adatoms on Si(111)-3$\times$2 surface. In addition, the second layer atom Si(X) obtain a little amount of charge (-0.15$e$ for Ca, -0.09$e$ for Sr, and -0.07$e$ for Ba) from AEM atoms, and the corresponding charge density is clearly larger than the neighboring Si(X$^{\prime}$) atom along the empty channel. The effective charge of Si(E) is -0.17$e$, -0.18$e$ and -0.21$e$ for the case of Ca, Sr and Ba atoms adsorption on Si(111)-3$\times$2 surface, respectively, which are also primarily supplied by AEM adatoms, and the variational tendency of charge transfer is opposite to other atoms in the top layer. As for the inner Si(C and D) and the second layer Si(Y) atoms, charege transfer are not obvious. However, it is important to note that our analysis of charge transfer only provide a qualitative picture for the core-level energy shifts. A quantitative analysis can be obtained by considering the "initial-state "and the "final-state" electron-core relaxation processes, as described in Refs. \cite{PRL1993-71-2338} and \cite{PRB1996-53-10942}.


\begin{figure}
\includegraphics[width=7.5cm]{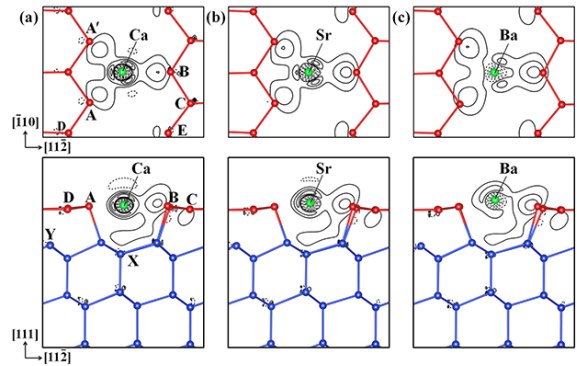}
\caption{(Color online) Total charge density difference plots for AEM adatoms on Si(111)-3$\times$2 surface. Solid (dashed) lines correspond to the gain (loss) of electrons. Red (blue) and green atoms represent Si and AEM atoms, respectively.}
\end{figure}

\subsection{Electronic band structures}

We calculate the electronic band structures based on hybrid density functional method (HSE06) \cite{HSE}. Due to the AEM/Si(111)-3$\times$2 surfaces have the same geometric structure \cite{EL2002-60-903,SS2002-512-269,PRB2002-66-165319,SS2016-653-222}, herein, we only show the band structure for Sr/Si(111)-3$\times$2 surface in Fig. 4, and the band structures for Ca and Ba have been provided in the supplementary materials. As show in Fig. 4(a), the band structure is plotted along [11$\bar{2}$] and [$\bar{1}$10] directions, the [11$\bar{2}$] direction corresponds to the $\bar{\Gamma}$-$\bar{C}$ direction, and the [$\bar{1}$10] direction corresponds to the $\bar{\Gamma}$-$\bar{A}$-$\bar{K}$-$\bar{C}$($\bar{M}$) direction. The symbols $\bar{M}$ and $\bar{K}$ are the symmetry points of (1$\times$1) surface Brillouin zone (SBZ), and $\bar{A}$ and $\bar{C}$ are the symmetry points of (3$\times$1) SBZ. The results show that the valence band maximum (VBM) is at $\bar{\Gamma}$ and the conduction band minimum (CBM) is along the $\bar{C}$-$\bar{\Gamma}$ direction, which indicate that Sr/Si(111)-3$\times$2 surface is an indirect band-gap semiconductor with a band gap of 1.44 eV. Like Sr, Ca and Ba adatoms induced Si(111)-3$\times$2 surfaces also present the semiconductive character with a 1.41 and 1.42 eV band gaps, respectively, and the difference is the CBM for Ca/Si(111)-3$\times$2 surface is at $\bar{C}$.  However, recently a band gap of 1.7 eV for Sr/Si(111)-3$\times$2 surface, reported by Du et al. \cite{SS2016-653-222}, was observed using scanning tunneling microscopy and scanning tunneling spectroscopy, which is larger than the band gap shown above. This phenomenon is, moreover, supported by the previous experimental findings \cite{PRB2004-69-125321,PRB2003-68-245312,EL2002-60-903,SS2002-512-269,PRB2001-64-165312,PRB2005-72-045310,PRB1997-55-6762,PRB2006-252-5292,PRB1994-50-1725,JPCM1998-10-3731,SS2009-603-727}. For example, the results measured by ARPES show that, in the valence band region, surface occupied states were blow the Si-bulk VBM for Ca/Si(111)-3$\times$2 \cite{PRB2004-69-125321,PRB2003-68-245312,EL2002-60-903,SS2002-512-269} and Ba/Si(111)-3$\times$2 \cite{PRB2001-64-165312} surfaces, and for the similar Yb/Si(111)-3$\times$2 \cite{PRB2006-252-5292} and Eu/Si(111)-3$\times$2 \cite{PRB2005-72-045310} surfaces. The similar findings were also observed on AM/Si(Ge)(111)-3$\times$1 surfaces \cite{PRB1997-55-6762,PRB1994-50-1725,JPCM1998-10-3731,SS2009-603-727}. These imply that the surface band gap is larger than the bulk band gap for AEM/Si(111)-3$\times$2 surfaces.

\begin{figure}
\includegraphics[width=7.5cm]{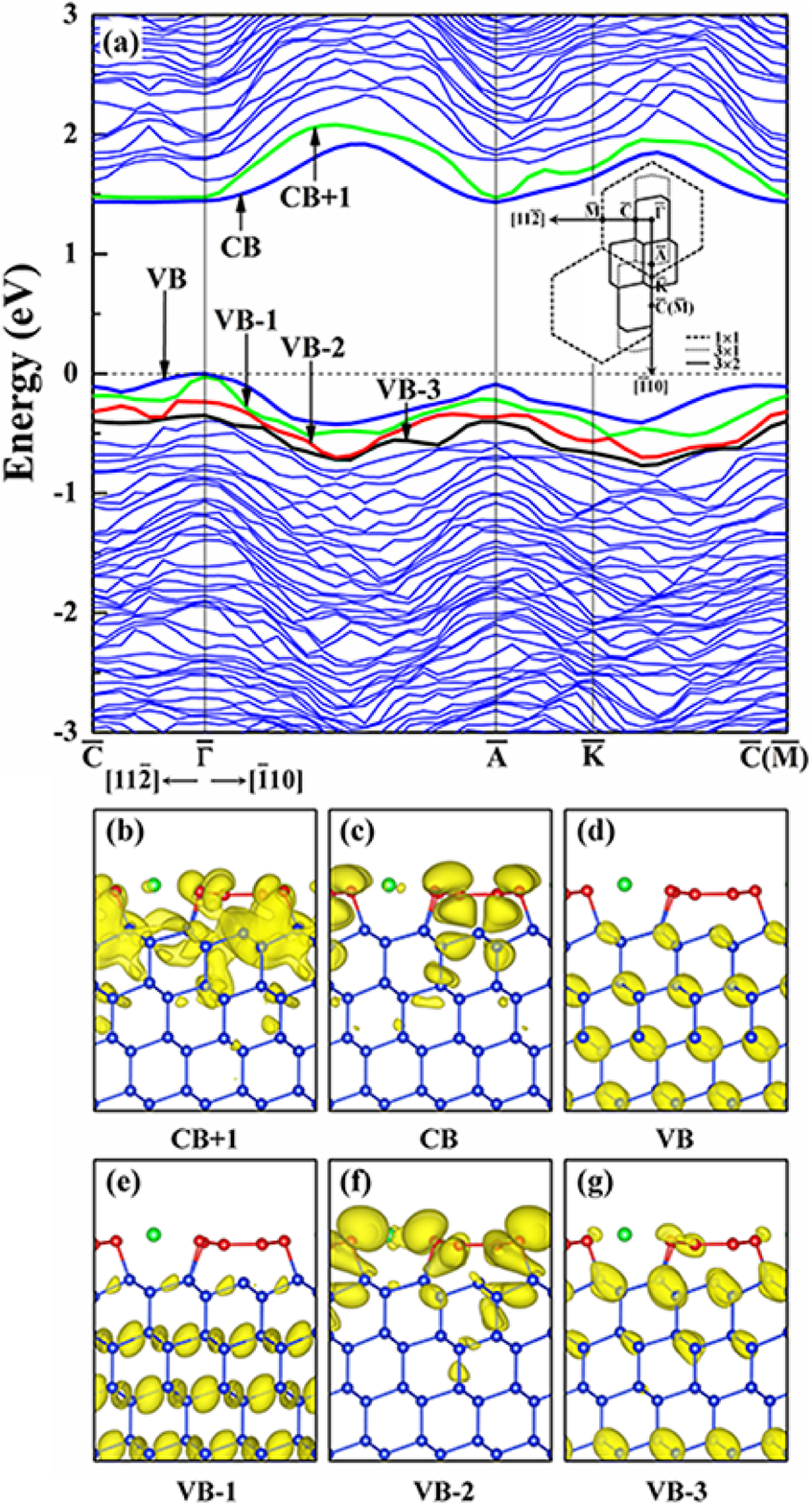}
\caption{(Color online) (a) Calculated electronic band structure of Sr/Si(111)-3$\times$2 surface using HSE06 functional. Horizonal dotted lines give the Fermi level that is set to zero. (b)-(g) depict the band-decomposed charge density distributions at $\bar{\Gamma}$: (b) the second lowest conduction band (CB+1), (c) the lowest conduction band (CB), (d) the highest valence band (VB), (e) the second highest valence band (VB-1), (f) the third highest valence band (VB-2), and (g) the fourth highest valence band (VB-3). The isovalue is 0.005 e/{\AA}$^{3}$. Red (blue) and green atoms represent Si and AEM atoms, respectively.}
\end{figure}

In order to investigate surface states, we further calculate the band-decomposed charge density distributions at $\bar{\Gamma}$ near Fermi level. As shown in Fig. 4(d) and (e), there are no contribution of charge from top layer atoms for the highest valence band (VB) and the second highest valence band (VB-1), and the almost electronic charge distribution are derived from bulk Si atoms, therefore VB and VB-1 are the bulk states. Different from VB and VB-1, the third and four highest valance band (VB-2 and VB-3) [see Fig. 4(f) and (g)] are the surface states because of the existence of charge contribution from surface atoms. The surface state VB-2 clearly represent the saturated dangling-bond states of the Si (A, B and E) atoms, and its charge distribution is primarily attribute to the the $d_{xy}$ orbital of Sr adatom and the $p_{x}$ orbital of surface Si(A,B and E) atoms. For surface state VB-3, the charge contribution from top layer atoms is apparently reduced, and charge distribution is mainly derived from second and deeper layer Si atoms. Similarly, the lowest conduction band (CB), plotted in Fig. 4(c), is also a surface state and has a $\pi$$^{*}$ antibonding character between $p_{z}$ orbital of the Si(C) and Si(D) atoms, similar to surface state S$_{4}$ of Na/Si(111)-3$\times$1 surface \cite{PRB1998-58-R13359}, $U_{bc}$ of Ti/Si(111)-3$\times$1 surface \cite{PRL1998-81-2296} and the S$_{6}$ of Yb/Si(111)-3$\times$2 surface \cite{SS2010-604-1899}. Thus, a true Si double bond is indeed exist between Si(C) and Si(D) atoms, which is primarily responsible for the stability of HCC model \cite{PRL1998-81-2296}. Compared to the CB, the charge of surface state CB-1 (the second lowest conduction band) is mainly distributed around the second layer Si atoms, while the contribution of top layer atoms is little.

All the above results indicate that, surface band gap indeed exists between CB and VB-2, and the corresponding value are 1.65, 1.68, and 1.66 eV for Ca, Sr and Ba, respectively, which is good consistent with the experimental finding (1.7 eV) for Sr/Si(111)-3$\times$2 surface \cite{SS2016-653-222}.

\section{Conclusion}

In conclusion, we have performed {\it ab initio} calculations on the structure stability and electronic properties of AEM(Ca, Sr Ba)/Si(111)-3$\times$2 surfaces. Adsorption energy and charge density difference calculations show the high structural stability due to the strong chemical bonding.  Analysis of electronic band structures and band-decomposed charge density distributions indicates that the highest occupation band of surface states is third valence band, and the lowest unoccupation band is first conduction band. As a result,  a larger surface band gap of 1.65$-$1.68 eV is obtained, in agreement with the recent experimental finding (1.7 eV) for Sr/Si(111)-3$\times$2 surface \cite{SS2016-653-222}.
These results reveal a natural explanation for the relevant experimental observation and stimulate further experimental and theoretical exploration on the surface science.

\section{ACKNOWLEDGMENTS*}
This study was supported by the National Natural Science Foundation of China (Grant No. 11274356, 11374341) and the Strategic Priority Research Program of the Chinese Academy of Sciences (Grant No. XDB07000000).
C.F.C. acknowledges support by DOE under Cooperative Agreement DE-NA0001982.


\end{document}